\documentclass[12pt]{iopart}
\usepackage{iopams}
\usepackage{graphicx}
\usepackage{color}

\begin{document}

\title{A lattice polymer study of DNA renaturation dynamics}

\author{A. Ferrantini$^1$, M. Baiesi$^{1,2}$, E. Carlon$^1$}
\address{$^1$ Institute for Theoretical Physics, K.U.Leuven, Celestijnenlaan 200D, B-3001 Leuven, Belgium}
\address{$^2$ Dipartimento di Fisica and sezione CNISM, Universit\`{a} di Padova, Padova, Italy}
\ead{alessandro@itf.fys.kuleuven.be}

\begin{abstract}
DNA renaturation is the recombination of two complementary single
strands to form a double helix. It is experimentally known that
renaturation proceeds through the formation of a double stranded nucleus
of several base pairs (the rate limiting step) followed by a much faster
zippering. We consider a lattice polymer model undergoing Rouse dynamics
and focus on the nucleation of two diffusing strands.  We study numerically
the dependence of various nucleation rates on the strand lengths and on an
additional local nucleation barrier.  When the local barrier is sufficiently
high, all renaturation rates considered scale with the length as predicted
by Kramers' rate theory and are also in agreement with experiments:
their scaling behavior is governed by exponents describing equilibrium
properties of polymers. When the local barrier is lowered renaturation
occurs in a regime of genuine non-equilibrium behavior and the scaling
deviates from the rate theory prediction.
\end{abstract}

\submitto{JSTAT}

\maketitle

\section{Introduction}
When a double-stranded DNA molecule is heated above some characteristic
temperature it undergoes a denaturation or melting transition and the
two strands separate \cite{wart85}.  The reverse process can also occur: two
complementary single stranded nucleic acids at temperatures below their
melting point bind to form a double-helix. This transition is known
as renaturation \cite{cant80}.

The first investigation of the DNA renaturation kinetics dates back to the work of
Wetmur and Davidson of the late sixties \cite{wetm68}. They experimentally
measured the renaturation rate $k_2$ and found it scales as a function
of $L$, the length of the strands,
\begin{equation}
k_2 \sim L^\alpha
\end{equation}
with $\alpha = 0.5$.
A recent detailed analysis of experimental data for DNAs from different
organisms, with different conditions of salt and temperature was performed
by Sikorav et al. \cite{siko09} who estimated $\alpha = 0.51(1)$.

The origin of the exponent $\alpha$ has been discussed in the literature
\cite{wetm68,schm72,siko09}. Renaturation is believed to proceed via
two steps: the formation of an active nucleus of a few base pairs,
which is followed by a rapid zippering until the double helix is
formed over its full length \cite{cant80}. The rate-limiting step is
the nucleus formation, therefore theoretical investigations of $k_2$
focus on the nucleation mechanism from the two separate strands.
It was first believed that $\alpha =1/2$ could be explained by the
theory of diffusion limited reactions \cite{schm72}.  This is however
at odds with more modern concepts of polymer physics, as explained in
Ref.~\cite{siko09}: the correct analysis of diffusion limited reaction
would imply a scaling of the type $k_2 \sim L$, which is not consistent
with the experiments. 
According to Sikorav et al.~\cite{siko09},
experimental data suggest 
that renaturation is an activated process. Using
Kramers' theory, in which the polymer dynamics is mapped into that of
an effective Brownian particle escaping a potential barrier, they find
that the rate of nucleation of two monomers scales as \cite{siko09}:

\begin{equation}
k_{\rm mon} \sim L^{\sigma_4} 
\label{k_mon}
\end{equation} 
where the exponent $\sigma_4$ is that associated to the equilibrium contact
probability through the middle monomer of the two renaturating polymers
(see next section). As nucleation can occur in any of the
$L$ monomers of the two approaching strands, one expects \cite{siko09}
the renaturation rate to scale as:
\begin{equation} 
k_2 \sim L k_{\rm mon} \sim L^{1+\sigma_4}.
\label{sigma4} 
\end{equation}
Using the numerical estimate $\sigma_4=-0.48$ \cite{hsu04} for
self-avoiding walks in three dimensions one gets $\alpha = 1 - 0.48 =
0.52$ in good agreement with experiments.

The argument leading to Eq.~(\ref{k_mon}) is based on a mapping of the DNA
renaturation problem onto that of a single Brownian particle crossing a
free energy barrier \cite{siko09}. The barrier has an entropic origin:
it is generated by excluded volume interactions of the two approaching
strands.  One applies then Kramers' rate theory, which states that
the transition rate is proportional to $e^{\Delta F/k_B T}$
for a free-energy barrier $\Delta F$ and a temperature $T$. 
For two polymers of length $L$ one has $\Delta F = k_B T \sigma_4 \log L$ (see Eq.~(\ref{free_en_barrier})),
from which one obtains  Eq.~(\ref{k_mon}).

Mapping the complex polymer dynamics into that of a single effective
Brownian particle performing a one dimensional motion is a huge
simplification of the problem. Moreover Kramers' approach assumes that
the polymer dynamics is sufficiently slow so that at any given time
the conformation of the polymer is well approximated by its equilibrium
distribution.

To gain more insights into the origin of the exponent describing the scaling behavior of the rate constant $k_2$ as observed in experiments, we performed numerical simulations of a coarse-grained model of DNA renaturation. The model is similar to those used in the past for the study of equilibrium properties of the DNA denaturation transition \cite{caus00,carl02,ever07}. It does not incorporate helical degrees of freedom of the double stranded DNA as the rate limiting step is nucleation, i.e. the first ``reactive'' contact between the approaching strands. The lattice nature of the model allows the simulation of long times and long polymers and, thus, an accurate determination of the scaling properties of the renaturation rates.

The dynamics of the DNA denaturation has attracted quite some attention in the recent literature
\cite{bar07,foge07,baie09}, but the reverse process of renaturation has
been less studied.  Our results show that the renaturation process has
a rich dynamics and that the scaling predicted by Kramers' rate theory
(Eqs. (\ref{k_mon}) and (\ref{sigma4})) is only valid in a specific
limiting case.

This paper is organized as follows: We start in the next section with a review of a few theoretical results about scaling properties of equilibrium free energies.
In Section~\ref{sec:model} we
introduce the model used in this study and in Section~\ref{sec:eq}
we present its equilibrium properties. The simulation
results for the rate constants for renaturation are presented
in Section~\ref{sec:rates}. Finally, results are discussed in
Section~\ref{sec:discussion}.

\section{Scaling properties of equilibrium free energies}
\label{app}

\begin{figure}[!tb]
\begin{center}
\includegraphics[height=10cm]{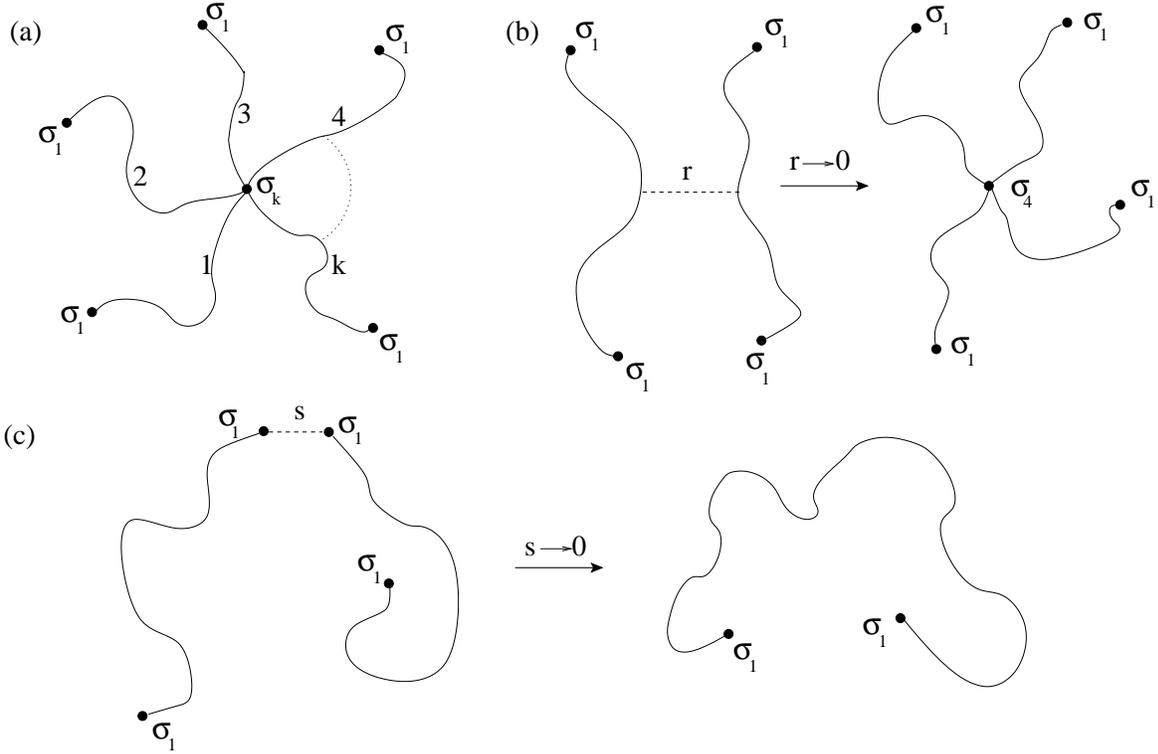}
\caption{(a) An example of star polymer with $k$ arms. There are $k$ vertices of order 1 and one of order $k$. (b) Nucleation of two polymers from the middle monomer. The contact leads to the formation of a star polymer with four outgoing legs. (c) Nucleation of two polymers from end monomers. The result after nucleation is only one polymer of double length with 2 vertices of order 1.
\label{sigma}}
\end{center}
\end{figure}

Before entering into the details of our simulations, we briefly recall the scaling properties of the free energy required to bring two self avoiding walks (SAWs) at a given fixed distance $r$ between their central monomers. Field theoretical models for swollen polymer systems predict that the canonical partition function of a star polymer with $k$ outgoing arms of length $L$ is given by \cite{dupl86}:
\begin{equation}
Z_L^{(k)} \sim \mu^{kL} L^{\sigma_k + k \sigma_1}
\label{part_Z}
\end{equation}
in the limit $L \to \infty$. In Eq.~(\ref{part_Z}) $\mu$ is a non-universal parameter (the connectivity constant), while $\sigma_k$'s are universal exponents associated to a vertex with $k$ outgoing legs. In a star there are $k$-ends (each contributing with an exponent $\sigma_1$) and one vertex with $k$ outgoing legs (contributing with an exponent $\sigma_k$), as shown in Fig.~\ref{sigma}~(a).The value of the exponents $\sigma_k$ is known from numerical simulations \cite{hsu04} and field theoretical computations using $4-\varepsilon$ expansions and resummation techniques \cite{scha92}.

Consider now two SAWs of length $L$, whose central monomers are separated
by a vector $\vec{r}$ from each other. The partition function is given by:
\begin{equation}
Z(\vec r) \sim \mu^{2L} L^{4 \sigma_1} 
g\left( \frac{\left|\vec r \right|}{L^\nu} \right)
\label{part_r}
\end{equation}
where $g(x)$ is a scaling function and $\nu$ the SAW metric exponent. When
the two polymers are sufficiently far from each other $r \gg L^\nu$,
or $x \to \infty$ one has $g(x) \to 1$, as the partition sum in
Eq.~(\ref{part_r}) should reduce to $(Z_{\rm SAW})^2$, 
where $Z_{\rm SAW} \sim \mu^L
L^{2 \sigma_1}$ is the partition function of a single self-avoiding walk.
For $r \ll L^\nu$, Eq.~(\ref{part_r}) should reduce to that of a 4-arm
star polymer $Z_{L/2}^{(4)}$ (see Fig.~\ref{sigma}~(b)), which imposes the following scaling behavior
in the limit $x \to 0$
\begin{equation}
g(x) \sim x^{-\sigma_4/\nu}.
\label{limit_gx}
\end{equation}
The numerical value for the exponent of three dimensional self-avoiding
walks are $\sigma_4 \simeq -0.48$ \cite{hsu04}, $\nu \simeq 0.588$ \cite{vand98}, 
therefore $-\sigma_4/\nu \simeq 0.82$. From the limiting behaviors for $| \vec r | \to \infty$
and $| \vec r | \to a$ (a microscopic distance) one obtains the
free energy difference between a configuration in which the two
central monomers are far apart and a configuration where the monomers
are at a distance $a$ as
\begin{equation}
\Delta F =  k_B T \sigma_4 \left(\log L - \frac 1 \nu \log a \right)
\label{free_en_barrier}
\end{equation}

One can repeat the same reasoning for contacts between end monomers (see Fig.~\ref{sigma}~(c)). Let $\vec s$ be the separation between two end points of the polymers. The partition function is
\begin{equation}
Z(\vec s) \sim \mu^{2L} L^{4 \sigma_1} 
h\left( \frac{\left|\vec s \right|}{L^\nu} \right)
\label{part_s}
\end{equation}
where again $h(x) \to 1$ for $x \to \infty$, while the short distance
limiting behavior is
\begin{equation}
h(x) \sim x^{2 \sigma_1/\nu}
\label{limit_hx}
\end{equation}
with $\sigma_1 \simeq 0.08$, hence $2\sigma_1/\nu \simeq 0.27$.

Similar scaling arguments were applied to the study of thermal denaturation of double stranded DNA \cite{kafr02}.

\section{Model}
\label{sec:model}

We consider two self-avoiding walks of length $L$ on a face centered cubic
(fcc) lattice, in which each site has $12$ nearest neighbors displaced by
$(\pm 1,\pm 1,0)$, $(\pm 1,0,\pm 1)$, and $(0,\pm 1,\pm 1)$.
A configuration of the two walks is specified by the position of
their monomers $\vec{r}_1 (i)$, $\vec{r}_2 (i)$, with $i=1, 2, \ldots, L$.
The two walks are also mutually avoiding, except at homologous monomers
which can overlap, i.e. $\vec{r}_1 (i) = \vec{r}_2 (j)$ is only allowed
if $i=j$. In the renaturated state the two walks overlap over their
full length, $\vec{r}_1 (i) = \vec{r}_2 (i)$ for all $i$.  The version
of this model on the simple cubic lattice was introduced by Causo et
al. \cite{caus00} for studying the thermodynamics of DNA denaturation, i.e. the unbinding of a double stranded polymer due to a temperature increase.

In our simulations the starting configuration consists of two separate strands placed at random in a cubic box with periodic boundary conditions. The walks are first equilibrated while keeping their central monomer fixed. At time $t=0$ the polymers begin their diffusive motion until nucleation, i.e. the contact between homologous monomers, occurs. At that point the simulation ends and the nucleation time is registered. The polymer dynamics consists of local corner-flip or end-flip moves that do not violate self-avoidance. The probability that a given monomer moves is taken proportional to the number of nearest neighbors that can be reached on a fcc lattice: we randomly attempt on average $11$ Monte Carlo moves for each end-monomer and $3$ moves for each other monomer.

We performed simulations for polymers of various lengths up to $L=193$, within a cubic simulation box with side $100$ (the lattice parameter being $a=\sqrt{2}$). For the longest polymers considered in this work, the gyration radius is $R_g \approx 12$, which is still much smaller than the box size, thus the simulations are performed in the diluted regime. We verified that the simulations reproduce the expected Rouse dynamics behavior $D \sim 1/L$ \cite{doi89, panj09} for a self avoiding walk. This model neglects hydrodynamics effects, which however are not expected to change the value of the exponents, at least in some regime (see Discussion).

The Monte Carlo dynamics satisfies detailed balance, therefore it would be possible also to infer the renaturation rate from an initial bound state and waiting until unbinding takes place. Here we prefer to use the direct simulation approach, as it is closer to the experimental setup.

\section{Equilibrium properties}
\label{sec:eq}

We focus first on the equilibrium probability distributions of two
strands to verify the scaling form of the partition functions derived in
Section 3. We ignore nucleation events and in a single long run we sample
the equilibrium configurations of the diffusing strands.  Let us denote
with $r$ the distance between the middle monomers of the two strands and
let $P(r)$ be the probability distribution function for $r$. If $r \gg
R_g$, with $R_g$ the radius of gyration of the two polymers, then $P(r) \sim n(r)$, where
$n(r)$ is the number of points at distance $r$ from a given point in an
fcc lattice. Asymptotically in $r$ we have $n(r) \sim r^2$, equivalent
to uniform probability in three dimensional space.

The probability of finding the two central monomers at distances $r
\lesssim R_g$ is suppressed due to mutual avoidance between the two
interpenetrating strands.  In order to get a good statistics on $P(r)$
for  $r \lesssim R_g$ we bias the dynamics so that configurations with
small $r$ are favored.  This biased dynamics works as follows.  A move
changing the distance between the middle monomers from $r$ to $r'$ is
accepted with probability $\min\{1, n/n'\}$, where $n$  and $n'$ are the
number of fcc lattice points at a distance $r$ and $r'$ from the origin,
respectively.  The biased dynamics produces a probability distribution
$P^*(r)$ which scales as $\displaystyle P^*(r) \sim \frac{P(r)}{r^2}$
in the limit of large $r$.
This construction is particularly useful because
we have $P^*(r)\sim g(r L^{-\nu})$, with $g(x)$ introduced in
Eqs.~(\ref{part_r})-(\ref{limit_gx}), that is, we can investigate directly
the scaling function $g(x)$.

\begin{figure}[!tb]
\begin{center}
\includegraphics[height=6.5cm]{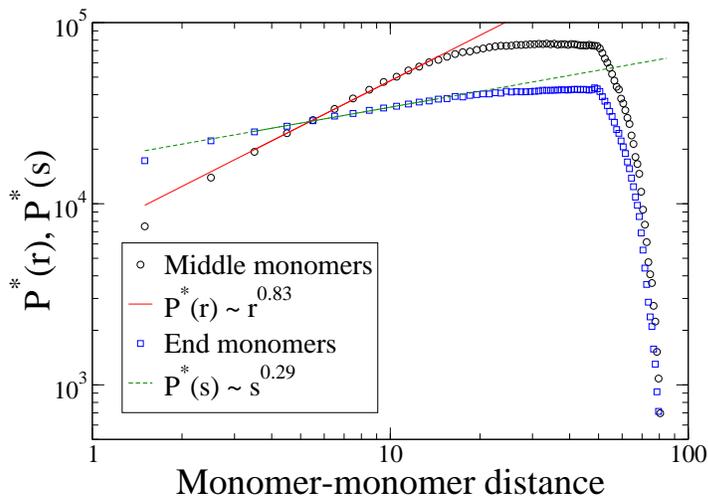}
\caption{Log-log plot of $P^*(r)$ and $P^*(s)$, the reweighted equilibrium distribution of distances between middle monomers (black circles) and end monomers (red squares) respectively. The polymers simulated consist both of 97
monomers. The decay at long distances is due to the finite size of the cubic box, while at short distances self-avoidance interactions play a central role. A fit of the data at short distances gives $\displaystyle P^*(r) \sim r^{0.83 \pm 0.05}$, and $\displaystyle P^*(s) \sim s^{0.29 \pm 0.05}$, both in agreement with the scaling form discussed in Eq.~(\ref{limit_gx}) and Eq.~(\ref{limit_hx}).\label{mm_distance}}
\end{center}
\end{figure}

A plot of a $P^*(r)$ for two strands with $L=97$ is shown in Fig.~\ref{mm_distance} (circles). For short distances the data reproduce the power-law behavior discussed in Eq.~(\ref{limit_gx}). For intermediate distances $P^*(r)$ is constant, a signature that the two polymers are sufficiently far apart to be considered non-interacting ($P(r) \sim r^2$), whereas for longer distances the probability decays because of the finite size of the simulation box.  

In Fig.~\ref{mm_distance} (squares) we also show a similar plot for $P^*(s)$, the reweighted distance between the end monomers. For the calculation we used again biased dynamics, where the bias is on the distance between two end monomers and it is constructed with similar rules as for the central monomers. The results are consistent with the power-law limiting behavior of Eq.~(\ref{limit_hx}).

\section{Renaturation rates from simulations}
\label{sec:rates}

We now turn to the renaturation dynamics. As mentioned in the Introduction, the rate limiting step in DNA renaturation is the formation of a nucleus of a few base pairs, from where a rapid zippering follows. In the simulation we focus on the time needed to reach a nucleation event and the simulation is stopped when nucleation occurs. The average nucleation time $\langle t_{\rm nucl.} \rangle$ is calculated from typically about $10^3$ independent realizations, i.e. every time randomly generating the starting configuration of the two strands. This translates into an average elapsed real time of about two weeks on an Intel(R) Core(TM)2 Quad CPU Q9650 $@$ 3.00GHz processor.
The nucleation rate is then obtained by:
\begin{equation}
k = \left\langle t_{\rm nucl.} \right\rangle ^{-1} .
\label{MC}
\end{equation}

\begin{figure}[!tb]
\begin{center}
\includegraphics[height=7cm]{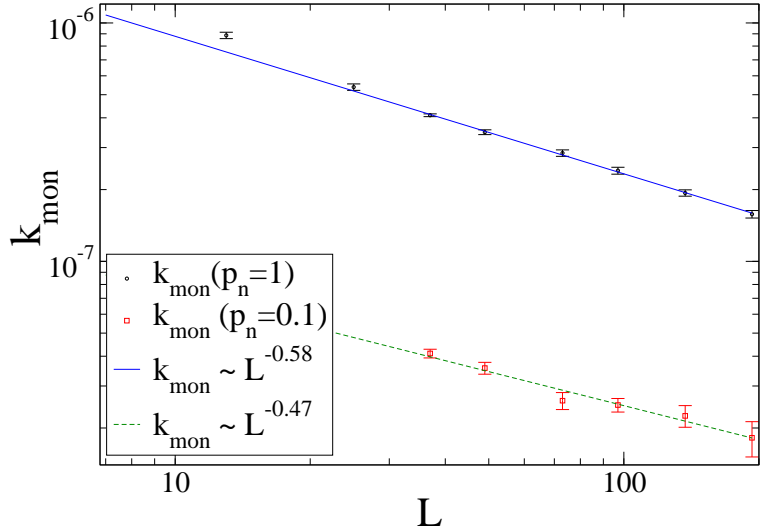}
\caption{Log-log plot of $k_{\rm mon}$ as a function of the polymer length
$L$. $k_{\rm mon}$ is obtained by means of Eq.~(\ref{MC}) from an average of
first contact times of middle monomers over different realizations with
the polymers being initially placed at random in the box (size=100). As
a result $k_{\rm mon} \sim L^{-0.58 \pm 0.01}$ for $p_n=1$ and $k_{\rm mon}
\sim L^{-0.47 \pm 0.04}$ for $p_n=0.1$. \label{alpha_mm}}
\end{center}
\end{figure}

It has been observed experimentally that the two strands must overcome
a free energy barrier of about $30 k_B$J$/$mol \cite{thro68} to form an
active nucleus and that the nucleus corresponds to a double helix of
about $15$ bases \cite{mann76,cant80}. Hence, the diffusion process
may bring frequently two complementary monomers in contact, but many
of these encounters are not ``productive'' as they do not lead to
the formation of an active nucleus from which zippering of the helix
would follow. We model this feature by introducing a local nucleation
probability: a move that would bring two homologous monomers 
in contact [say $\vec{r}_1 (i) = \vec{r}_2 (i)$] is
accepted with probability $p_n$, otherwise it is rejected and the simulation
continues. Hence, low values of $p_n$ correspond to high nucleation
barriers and consequently long simulation times. To avoid confusion we will
refer to the latter as a local barrier. This is not related to the free energy
barrier $\Delta F$ originating from the mutual self avoidance of the
renaturating strands discussed in the Introduction and given in Eq.~(\ref{free_en_barrier}). We will show below that the
estimates of the scaling exponents depend on $p_n$.

Nucleation times corresponding to different values of $p_n$ can be
obtained during the same run. Consider a set of nucleation probability
values $p_n^{(1)} < p_n^{(2)} < \ldots < p_n^{(k)}$, each representing a
different system. At an attempt of a nucleation step we generate a random
number $r$ uniformly distributed in the unit interval, and we consider as nucleated all systems $q$ for
which $r<p_n^{(q)}$. Their nucleation time is added to the corresponding
statistics, and the simulation continues in order to sample the nucleation
times of the remaining systems, until nucleation for the smallest
value $p_n^{(1)}$ has been reached.

\subsection{Rate of middle monomers nucleation}

We consider first the nucleation between middle monomers. In this case
the nucleation time is defined as the time for first contact between
the two central monomers $i = (L-1)/2$ (for convenience we considered
polymers with an odd number of monomers so that there is a unique middle
monomer). In the course of the simulation two other analogous monomers
may come in contact, say $\vec{r_1}(k) = \vec{r_2}(k)$ with $k \neq
(L-1)/2$, but the event is ignored and the simulation continues.

In Fig.~\ref{alpha_mm} we show our numerical results using nucleation
probabilities $p_n = 1$ and $p_n = 0.1$ and for polymers of length
up to $L = 193$.  For each length we typically sample about $10^3$
independent nucleation events.  Both nucleation probabilities give a
nucleation rate decaying as a power-law as a function of the length
$L$. However the exponent depends on $p_n$.  Our estimates, which are
based on linear fits for $L \geq 30$, yield $k_{\rm mon} \sim L^{-0.58
\pm 0.01}$ for $p_n = 1$ (corresponding to the case of no local barrier)
and $k_{\rm mon} \sim L^{-0.47 \pm 0.04}$ for $p_n = 0.1$.

Kramers' theory (Eq.~(\ref{k_mon})) \cite{siko09} predicts for $k_{\rm
mon}$ a decay governed by an exponent $\sigma_4 = -0.48$. We note that our
numerical data are in agreement with that for $p_n = 0.1$. The results
for $p_n = 1$ are however not compatible with the Kramers' theory. 

\subsection{Rate of polymer nucleation}


\begin{figure}[!tb]
\begin{center}
\includegraphics[height=7.5cm]{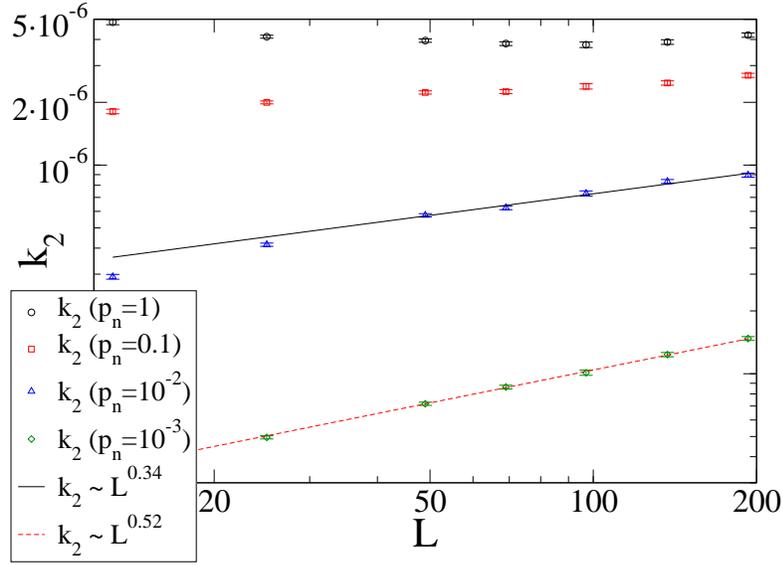}
\caption{Log-log plot of $k_2$ as a function of the polymer length $L$ for different contact probabilities $p_n$. The points are obtained by taking the inverse of the average time until first contact. Four series are shown corresponding to $p_n=1$, $0.1$, $0.01$, $0.001$. The fit for $p_n=0.001$, $L>30$, gives $k_2 \sim L^{0.52 \pm 0.01}$.}
\label{alpha}
\end{center}
\end{figure}


We consider next the nucleation rate $k_2$ of the whole polymer. We
define the nucleation time $t_{\rm nucl}$ as the first contact time
between any of the $L$ complementary monomers. Figure~\ref{alpha} shows plots of $k_2$
as a function of $L$ for different values of the nucleation probability
($p_n = 1, 0.1, 0.01, 0.001$).  In absence of a local nucleation barrier
($p_n = 1$) we find a non-monotonic behavior for $k_2$. It decreases for
$L \lesssim 50$, while it increases for larger $L$. As the value
of $p_n$ is lowered, $k_2$ becomes monotonic in $L$. From a fit of the
data for $L \geq 30$ we obtain the estimates $\alpha = 0.34 \pm 0.01$
for $p_n = 0.01$, while $\alpha = 0.52 \pm 0.01$ for $p_n = 0.001$.
Again, the exponent is consistent with that predicted by the Kramers'
theory (Eq.~(\ref{sigma4})) only for sufficiently low $p_n$.

\begin{figure}[!tb]
\begin{center}
\includegraphics[height=7.5cm]{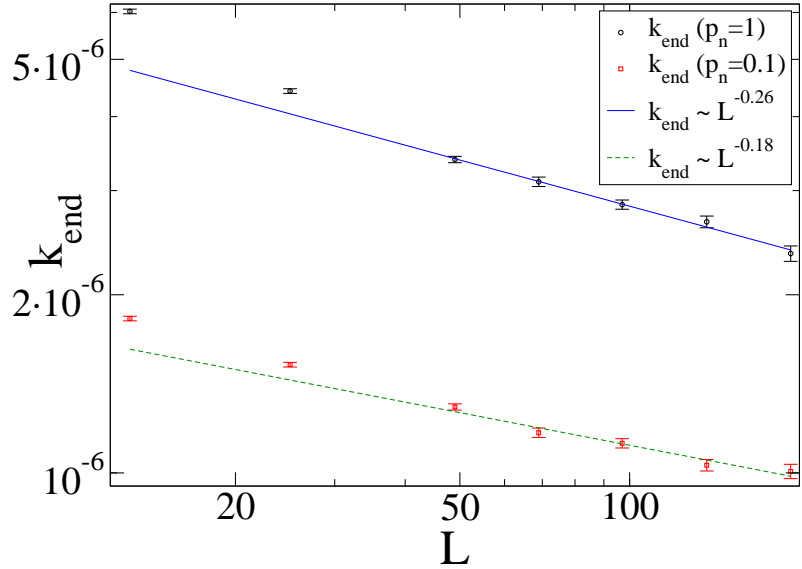}
\caption{Log-log plot of $k_{\rm end}$ as a function of the polymer length $L$. The points are obtained by taking the inverse of the average time until (on average) the ${\rm 10^{th}}$ contact between end monomers. The fit for $L \gtrsim 30$ gives $k_{\rm end} \sim L^{-0.26 \pm 0.01}$ for $p_n=1$ and $k_{\rm end} \sim L^{-0.18 \pm 0.02}$ for $p_n=0.1$.
\label{alpha_em}}
\end{center}
\end{figure}

\subsection{Rate of end monomers nucleation}


We finally studied the renaturation dynamics for end monomers. In
this case we define the nucleation time as the time needed to
nucleate the monomers at one edge of the strand, at position $i=1$.
Figure~\ref{alpha_mm} shows the data of $k_{\rm end}$ as a function of $L$
for two different values of the local nucleation probability.  For $p_n=1$
we find as best fit to the data $k_{\rm end} \sim L^{-0.26 \pm 0.01}$,
while $k_{\rm end} \sim L^{-0.18 \pm 0.02}$ for $p_n = 0.1$.
We also considered a smaller value of $p_n = 0.01$ and found 
$k_{\rm end} \sim L^{-0.15 \pm 0.02}$ (data not shown). This suggests 
that a limiting value of the exponent has been reached. Kramers' theory
would predict a scaling for the renaturation rate of the end monomers
as $k_{\rm end} \sim L^{-2 \sigma_1} = L^{-0.16}$ (see Eq.~\ref{sigma4}).
Again for sufficiently low $p_n$ (high local nucleation barriers) 
the simulations results are in agreement with Kramers' theory.


\section{Discussion}
\label{sec:discussion}

In this paper we investigated a simple polymer lattice model of DNA
renaturation. We focused on the scaling of the renaturation rate as a
function of the polymer length. Three different rates were considered: the
rates of renaturation for middle monomers ($k_{mon}$), for end monomers
($k_{end}$) and for the whole strands ($k_2$). We also introduced a
local nucleation probability $p_n$ which is associated to an additional
{\it local} free energy barrier. In DNA this barrier corresponds to the
free energy one has to overcome to form an active nucleus of about 10-15
bases from where a rapid zippering starts.

\begin{figure}[!tb]
\begin{center}
\includegraphics[height=5.5cm]{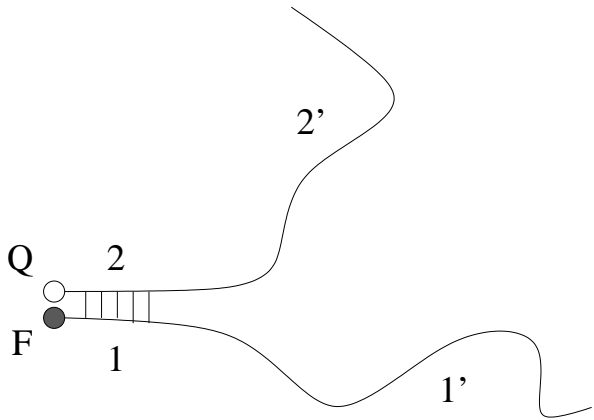}
\caption{Schematic representation of the suggested experiment to measure
the nucleation rate of end-monomers $k_{end}$. The strands have sequences
of length $l$ that are complementary at one end (labeled as 1 and 2).
The rest of the strands is composed by a non-complementary
$L-l$ stretch (such as poly(A) stretches) so that nucleation cannot
occur in this part. The end-renaturation dynamics can be detected by
measuring the light emitted by a couple of quencher-fluorophore, as
used in molecular beacons. 
\label{beacon}}
\end{center}
\end{figure}

The simulations show that if the local barrier is sufficiently high
($p_n \ll 1$) the exponents governing the scaling behavior of
$k_{mon}$, $k_{end}$ and $k_2$ are all in agreement with Kramers' rate
theory. An increase of the local barrier corresponds to an increase
of the simulation time since the polymers attempt an increasing
number of unsuccessful nucleation events. The process is slow,
therefore the probability distribution of finding two monomers at a
given distance from each other approaches its equilibrium value. In
this regime Kramers' theory is expected to be valid. In addition, the
whole strand renaturation rate $k_2$ scales with an exponent consistent
with experimental data \cite{siko09}, which suggests that experimental
conditions correspond to the high local barrier regime. Experiments so
far focused on the rate $k_2$, which is usually measured by monitoring
the time dependence of the UV absorption spectrum. We argued, and
verified in simulations in the high local barrier regime ($p_n \ll 1$),
that the rate for end monomer nucleation scales as $k_{\rm end} \sim L^{-0.18}$. By
appropriately engineering the sequences of the renaturating strands
one could mesure $k_{\rm end}$ in experiments. For this purpose it is
sufficient to consider two sequences which are complementary in $l$ end
bases and non-complementary for the remaining $L-l$ bases (see Fig.~\ref{beacon}). As
hybridization can only occur at the edges, the dynamics is governed by
$k_{end}$. The detection can be done by a quencher-fluorophore pair as
used in molecular beacons \cite{tyag96,ambj06}. The scaling of $k_{end}$
is determined by increasing $L$ while keeping $l$ fixed. A modification
of the sequence architecture, e.g. using complementary stretches in the
middle of the sequence, would allow one to determine the renaturation
rate for a single monomer $k_{mon}$.

We also performed simulations in the low local barrier regime ($p_n \to
1$). In this case the exponents for $k_{mon}$, $k_{end}$ and $k_2$ deviate
from Kramers' rate theory predictions. The renaturation is ``fast''
so that two complementary monomers in the approaching strands do not
feel additional local barriers. In this regime it is natural to expect
that at least one of the hypothesis behind Kramers' theory breaks down,
namely the assumption that the two approaching strands can be considered
as being in quasi-equilibrium conformations. It would be tempting to
think that renaturation becomes a diffusion-limited reaction in this
regime, but this is actually not true as there is still a free energy
barrier (Eq.~(\ref{free_en_barrier})) generated by the self-avoidance
interactions between the strands. A better quantitative understanding of
this low local barrier regime remains challenging. We also expect that
the exponents in this fast renaturation regime would be influenced by
hydrodynamics effects which are not built in in our simulations. This
is an important point to bear in mind in view of possible comparison
with experimental data. Hydrodynamics should not influence however the
$p_n \ll 1$ regime as renaturation is slow and occurs between strands
in quasi-equilibrium regime.

Another remark concerns the limit of validity of Eq.~(\ref{sigma4}). As
the rate $k_2$ grows with $L$ (due to the increasing number of nucleation
sites as $L$ grows) we expect a deviation from Eq.~(\ref{sigma4}) at some
length $L^*$ beyond which renaturation becomes ``too fast'' in the sense
discussed above: in this regime conformations of the approaching polymers
are far from equilibrium. The scaling of $k_2$ as described by Kramers'
theory should hold for $L \lesssim L^*$.

In conclusion, our simulations of renaturation show that the process
has a very rich dynamical behavior. A key quantity appears to be the
local free energy barrier for monomer-monomer nucleation. Different
regimes appear in the limit of high barrier and of low barrier. The former
regime is well-understood within the framework of Kramers' rate theory,
while a better analytical understanding of the latter one remains
challenging.

\section*{Acknowledgements:}
We are grateful to G.T. Barkema, H. Orland, J.-L. Sikorav  and
C. Vanderzande for interesting discussions.

\section*{References}


\end{document}